\documentstyle[festkoer,psfig]{vieweg}

\author{G\"otz S. Uhrig}

\Kurzautor{Uhrig}

\title{Modulated Phases in Spin-Peierls Systems}

\Kurztitel{Modulated Phases in Spin-Peierls Systems}

\Adresse{Institut f\"ur Theoretische Physik, Universit\"at K\"oln,
Z\"ulpicherstra\ss{}e 77, 50937 K\"oln}

\begin{document}

\Titel

\begin{abstract}
Lattice modulations in the high magnetic field phase and
close to impurities in spin-Peierls systems are considered
and compared to experiment. Necessary extensions of existing theories
are proposed. The influence of zero-point fluctuations 
on magnetic amplitudes is shown.
\end{abstract}

The discovery of the first inorganic spin-Peierls
substance CuGeO$_3$ by Hase {\it et al.} \cite{hase93a} renewed
the interest in the phases of spin-Peierls substances
(for reviews, see \cite{bray83}).
The constituting element of these substances are
quasi one-dimensional spin chains coupled to the lattice
degrees of freedom.
The incommensurably modulated (I) phase is particularly
interesting since this phase is the most complex one from
the theoretical and from the experimental point of view.
Already 20 years ago much work was devoted to it
(e.g. [3-9]
\nocite{brazo80a,nakan80,merts81,horov81,buzdi83a,fujit84,hijma85}
).
Yet, detailed experimental investigations of the nature of
this phase were not possible at that time.
This has changed for CuGeO$_3$.

X-ray experiments in the I phase permitted  to detect the
incommensurability of the lattice in $k$-space \cite{kiryu95}.
The structure
of the lattice modulation was investigated by
measuring the intensity of the third harmonic \cite{kiryu95}.
The distribution of local magnetizations 
in CuGeO$_3$ was measured by NMR \cite{fagot96}. 
Recently, the shape and the
amplitude of the magnetic part of a soliton were 
deduced \cite{fagot96,horva99}.

Here we will review several aspects of the
theoretical description of the I phase. Special
emphasis is put on the discrepancies between the so far existing
theories and the experimental findings and to resolving these 
discrepancies  \cite{uhrig99b}.

The term ``soliton'' will be used for the combination of a zero 
in the modulated
distortion {\em and} the concomitant localized spinon 
\cite{nakan80,uhrig99a}. The distortive
soliton width $\xi_{\rm d}$ is the width of the kink-like modulation.
 The magnetic soliton width $\xi_{\rm m}$ is the spatial width of the local
magnetizations \cite{nakan80,schon98,uhrig99a}. The incommensurate
modulation in the I phase is viewed as an equidistant array (lattice)
of solitons.

\begin{figure}
\hfill\psfig{figure=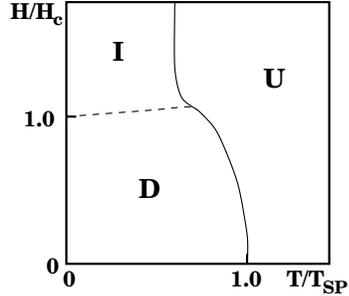,width=4.5cm}
\caption[fig1]{Generic phase diagram of a
spin-Peierls system. U: uniform phase;
D: dimerized phase; I: incommensurate phase.
Solid line:  second order transition;
dashed line: weak first or second order.}

\label{phasdia}

\end{figure}

The schematic phase diagram in Fig.~\ref{phasdia} shall serve
as a guide for the following discussion.
For the sake of concreteness we consider the hamiltonian
\begin{equation}
\label{hamilton}
H = \sum_iJ\left[(1+\delta_i){\bf S}_{i+1}{\bf S}_{i} +
  \alpha {\bf S}_{i+1}{\bf S}_{i-1}
\right] + \frac{K}{2} \delta_i^2
\end{equation}
which neglects the phonon dynamics (adiabatic treatment) since
the distortions $\delta_i$ are numbers  of which the  average is zero. 
The nearest neighbour coupling $J$ is modulated by the
$\delta_i$. Frustrating next-nearest neighbour
coupling $\alpha J$ is also considered. The spring constant $K$
measures the elastic energy of a given distortion.
Without magnetic field the spin chains 
are very susceptible towards dimerization $\delta_i=(-1)^i\delta$.
 The energy
lowering due to dimerization is anomalously large $\Delta E
\propto -\delta^{4/3}$ overcompensating the energy loss
in elastic energy $\propto \delta^2$ \cite{cross79}. 

Applying a magnetic field  to the dimerized (D) phase of a 
spin-Peierls system ($T$ small) leaves the D phase
unchanged until the critical field $H_c$ is reached.
The stability stems from the singlet-triplet gap $\Delta_{\rm trip}$.
Hence an upper bound for $H_c$ is $g\mu_{\rm B}H_c = \Delta_{\rm trip}$.
In fact, the critical field $H_c$ is smaller. Its Zeeman energy represents
only 80\% of the  gap $\Delta_{\rm trip}$
because  the modulation  also changes
above the critical field. The system
enters the incommensurably modulated phase.

\section{Sinusoidal Modulation} 
Elastic X-ray scattering confirmed \cite{kiryu95}
that the distortion in the I phase is incommensurate. The deviation
from dimerization $|q-\pi|$ increases with increasing magnetic field.
This can be illustrated  in the unfrustrated XY model which corresponds
via the Jordan-Wigner transformation to  free fermions.
The susceptibility towards distortion
becomes maximum at $q=2k_{\rm F}$ since $q=2k_{\rm F}$ allows
to create particle-hole pairs of vanishing energy. A distortion
with $q=2k_{\rm F}= 2\pi m + \pi$ is formed where  $m$ is the magnetization.
 Since this distortion couples the degenerate
states at $-k_{\rm F}$ and at $k_{\rm F}$ a gap appears there.
 Let us now gradually increase the interaction
corresponding to the $S^z_iS^z_{i+1}$ terms at fixed particle number. 
For continuity no state can cross the gap such that the picture of a
distortion at $2k_{\rm F}$ remains valid \cite{mulle81,uhrig98a}.
So we have
\begin{equation}
\label{qm-bedg}
|q-\pi| = 2\pi m \ .
\end{equation}
This relation is confirmed numerically \cite{schon98}.

Since experimentally it turned out that higher harmonics of the
distortion are considerably suppressed it is plausible to start with
($q$ given by (\ref{qm-bedg}) \cite{uhrig98a})
\begin{equation}
\label{sinus}
H = J \sum_i[1-\delta\cos(qr_i)] {\bf S}_i {\bf S}_{i+1} \ .
\end{equation}

The distribution of the local magnetizations $m_i=\langle S^z_i\rangle$
is found from NMR experiments \cite{fagot96,horva99}.
With some success, the experimental data were compared to a continuum
theory \cite{fujit84}. This theory, however, is based on a Hartree-Fock
treatment where all Hartree and Fock terms are  spatially constant.
In Ref. \cite{uhrig98a} it is shown that this is a too crude approximation
reducing the physics to the one of a XY chain. The antiferromagnetic
correlations found in this way are much smaller than those of an
isotropic XYZ chain. This conclusion is corroborated by several works
\cite{feigu97,forst98,schon98}. But the spin isotropy of
cuprates can hardly be questioned. To account for the smaller amplitudes
it is proposed \cite{uhrig98a,uhrig99b} that experimentally only an 
effective magnetization $m_i^{\rm eff}$ is seen which is 
an average
\begin{equation}
\label{mittel}
m^{\rm eff}_i = (1-2\gamma)m_i + \gamma(m_{i-1}+m_{i+1}) \ .
\end{equation}
The results for $\gamma=0.2$ agree well with
experiment \cite{uhrig98a}. The microscopic origin of the average
is discussed in Sect.~4.

The reason for strong local magnetizations around the zeros of the
modulation (cf. Fig.~\ref{inkomm}) is found in the localization of a spinon.
Each zero binds exactly one spinon \cite{uhrig99a}. Summing the $m_i$ 
around a magnetization maximum yields 1/2.

The order of the transition D  $\to$  I can be determined by investigating
the ground state energy $E(m)$ as function of the average magnetization
$m$ \cite{schon98}. By means of a Legendre transformation
$\tilde E(h)=E(m) - h m$, one obtains the dependence of ground state
energy  $\tilde E(h)$ on the magnetic field $h=g\mu_{\rm B} H$.
It is found that a discontinuous jump for $m\to0$ occurs. This
implies that the transition D $\to$ I for  fixed sinusoidal modulation is
of first order.  The mean square of $\cos(qr_i)$ jumps discontinuously 
from 1 to 1/2 if $q$ deviates infinitesimaly from $\pi$ \cite{schon98}
since the $r_i$ are summed over integer values only.

Experimentally, however, the observed first order jumps are much lower
than those found for fixed sinusoidal modulation 
\cite{palme96a,kiryu95,fagot96,ammer97,loren97a}.

\section{Adaptive Modulation}
Since it was stated above that sinusoidal modulation alone does
not account for the weak first order D  $\to$  I transition we turn to the
full minimization of the ground state energy of (\ref{hamilton}).
Derivation with respect to $\delta_i$ yields
\begin{equation}
 0 = \langle{\bf S}_{i+1}{\bf S}_{i}\rangle - 
\langle\langle{\bf S}_{j+1}{\bf S}_{j}\rangle\rangle +K\delta_i\ ,
\label{bondmin}
\end{equation}
where  $\langle\langle\cdot \rangle\rangle$ stands for
the expectation value and the average along the chain. The
double-bracketed term accounts for
the constraint of the vanishing average of the $\delta_i$.
The minimization is done iteratively \cite{feigu97,schon98,uhrig99b}.

\begin{figure}
\vspace*{3mm}
\hfill\psfig{figure=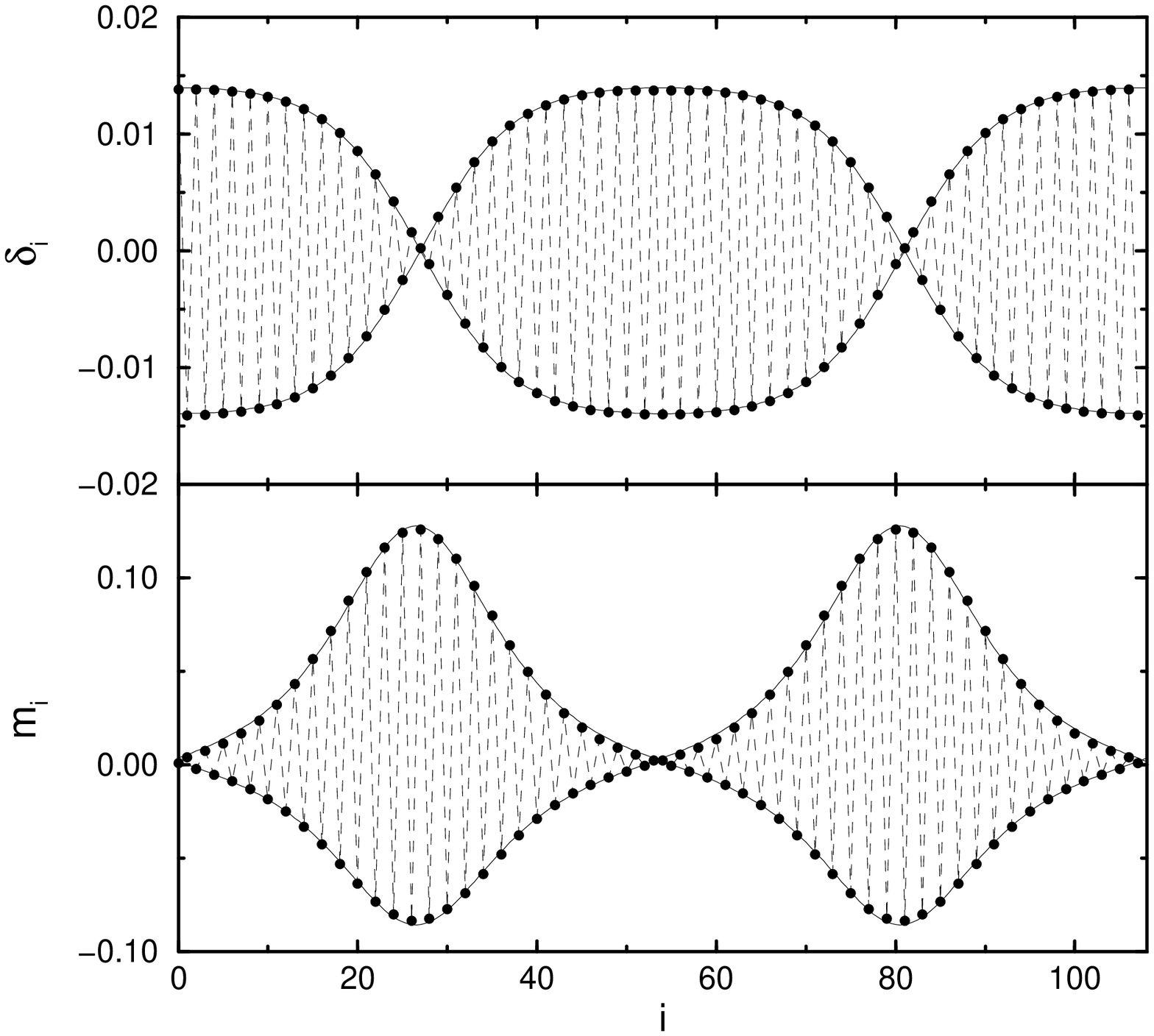,width=6.5cm}
\vspace*{-3mm}
\caption[fig2]{Symbols: DMRG-result at $K=18J$, $\alpha=0.35$.
Upper panel: local distortions;
 solid line: from
  Eq. (\protect\ref{fit1b}) with $\delta= 0.014$, $k_{\rm d}=0.959$,
 $\xi_{\rm d} = 10.5$.
Lower panel: local magnetizations;  solid line: 
 Eq. (\protect\ref{fit1a}) with $W=0.21 $, $R=5.0$, $k_{\rm m}=0.992$, 
$\xi_{\rm m} = 7.9$.}

\label{inkomm}

\end{figure}

The generic result is depicted in Fig.~\ref{inkomm}.
The local magnetizations do not display major differences to the 
results for sinusoidal modulation in \cite{uhrig98a} because 
the spinon localization is in essence determined only by the slope with
which the modulation vanishes.
Note that the envelope of $m_i$ is proportional to the probability
of finding a spinon at that site \cite{uhrig99a}. 
Hence, the magnetic part of the soliton displays localization as for 
sinusoidal modulation.
For the distortions the relation (\ref{bondmin}) implies that the 
deviations from constantly alternating dimerization are also localized.

The distortion belonging to an isolated soliton is a kink, i.e.
the distortion between two solitons resembles the one in the
D phase. This implies
a crucial advantage over the sinusoidal modulation. For kink-like
solitons the reduction
of the mean square distortion is proportional to the soliton number.
Hence, a low soliton concentration leads only to a small change of the
energy such that $E(m)$ is continuous (in the sense of Lipschitz) on
$m\to0$ \cite{schon98}.

The investigations in Ref. \cite{schon98}
of the model (\ref{hamilton}) yielded a continuous
phase transition even though the magnetization grows very quickly
above the critical field $m\propto -1/\ln(H-H_c)$.
Most of the results of the continuum theories comply also with a
phase transition of second order 
\cite{brazo80a,merts81,buzdi83a,fujit84}.
Solely Horovitz mentions the possibility of soliton attraction
in an early work \cite{horov81} implying a first order
transition. Buzdin {\it et al.} expect a first order transition at $T>0$
 \cite{buzdi83a}.

In fact, the details of the models matter.
Cross \cite{cross79b} argued already 
that an elastic energy {\it with} dispersion $K(q)$ being minimum at
 $q=\pi$ leads to a first order transition. The positive curvature
of  $K(q)$ around $q=\pi$
suppresses higher harmonics in the distortion. Hence, sinusoidal
modulation is favoured. The concomitant concavity in $E(m)$ at low
magnetization $m$ implies phase separation via the Maxwell construction.
So the phase transition is first order \cite{schon98}.
This finding is in accordance with the conclusion from a phenomenological
Ginzburg-Landau description \cite{bhatt98} that the
D $\to$  I transition is generically of first order.

At the transition the distance between two solitons is
rather large so that the difference between sinusoidal and adaptive modulation
matters most. At higher soliton concentrations the adaptive modulation
becomes more and more sinusoidal but with a concentration dependent amplitude.
 The mean square distortion could be determined from the elastic lattice
constants. These experimental results agree well with the 
predictions based on the model (\ref{hamilton}), see \cite{loren98}.

The continuum theories applying to the isotropic 
Heisenberg chain \cite{nakan80,zang97} provide the following
results (details in Ref. \cite{uhrig99b}; 
${\rm sn}, {\rm cn}, {\rm dn}$: elliptic Jacobi functions)
\begin{eqnarray}
\label{fit1a}
m_i &=& \frac{W}{2}\left\{ \frac{1}{R} 
{\rm dn}\left(\frac{r_i}{k_{\rm m}\xi_{\rm m}},k_{\rm m}\right)
+(-1)^i {\rm cn}\left(\frac{r_i}{k_{\rm m}\xi_{\rm m}},k_{\rm m}\right)\right\}\\
\delta_i &=& (-1)^i \delta\ 
{\rm sn}\left(\frac{r_i}{k_{\rm d}\xi_{\rm d}},k_{\rm d}\right)
\label{fit1b}
\\
\label{zus1}
\mbox{with}\qquad
\xi&:= &\xi_{\rm m} \, =\,  \xi_{\rm d} \quad \Leftrightarrow \quad
k \, :=\, k_{\rm m} \, =\, k_{\rm d}\\ \label{zus2}
1&=& 4 m k_{\rm m/d} K(k_{\rm m/d} )\xi_{\rm m/d} \\
 \label{zus3}
1&=& \pi k_{\rm m} \xi_{\rm m}\frac{W}{R}\ .
\end{eqnarray}

The fits in Fig.~\ref{inkomm} are based on Eqs. (\ref{fit1a},\ref{fit1b}).
Identity (\ref{zus1}) is {\it not} complied with, see Sect.~3. 
Otherwise no agreement
would be obtained. Relation (\ref{zus2}) is imposed on the fits whereas
Eq. (\ref{zus3}) serves as  check. It is fulfilled within 4\%. 
The fact that $\xi_{\rm d}/\xi_{\rm m}\approx 1.33$ is considerably
above unity complies nicely with the experimental findings. Elastic
 X-ray scattering \cite{kiryu95} found $\xi_{\rm d}=13.6\pm 0.3$ while
the NMR investigations provide $\xi_{\rm m}\approx 10$
\cite{horva99} close to the transition.

There is also another way to introduce
solitons in a spin-Peierls system than the application of magnetic field.
Doping non-magnetic  impurities in a spin-Peierls systems
cuts the infinite chains into finite chain segments \cite{khoms96}.
In a number of works
 \cite{marti96b} it has been shown that each
impurity frees one  spinon which is situated
either before or after the impurity on the chain.
Assuming a fixed dimerization it is easy to see that
the spinon is bound to its generating impurity \cite{uhrig99a}
in accordance with experimental
results \cite{hassa98,els98}.

But in a spin-Peierls system the change of the modulation has
to be taken into account, too. This is done by introducing
\begin{equation}
\label{stoerkopp}
H = J\sum_{i\ge 0}\left[
(1+\delta_i){\bf S}_i{\bf S}_{i+1} +\alpha {\bf S}_i{\bf S}_{i+2}+
\frac{K}{2} \delta_i^2 + f\delta_i(-1)^i\delta_{\rm bulk}
\right]\ ,
\end{equation}
such that the impurity is at site -1. The important amendment
compared to $H$ in Eq. (\ref{hamilton}) is the last term.
If the spinon moves away from the impurity the distortion
pattern is changed between impurity and spinon.
Due to an {\em elastic} interchain interaction (parametrized
by $f$)
a coherent distortion pattern throughout the whole three-dimensional
system is preferred. A deviation from this pattern  is
energetically unfavourable. So one is led to include the last
term in Eq. (\ref{stoerkopp}) assuming that the adjacent chains
are dimerized as in the unperturbed D phase. The relation  
$K=K_0+f$ ensures the consistency of the distortion 
amplitude with its bulk value. 

\begin{figure}
\hfill\psfig{figure=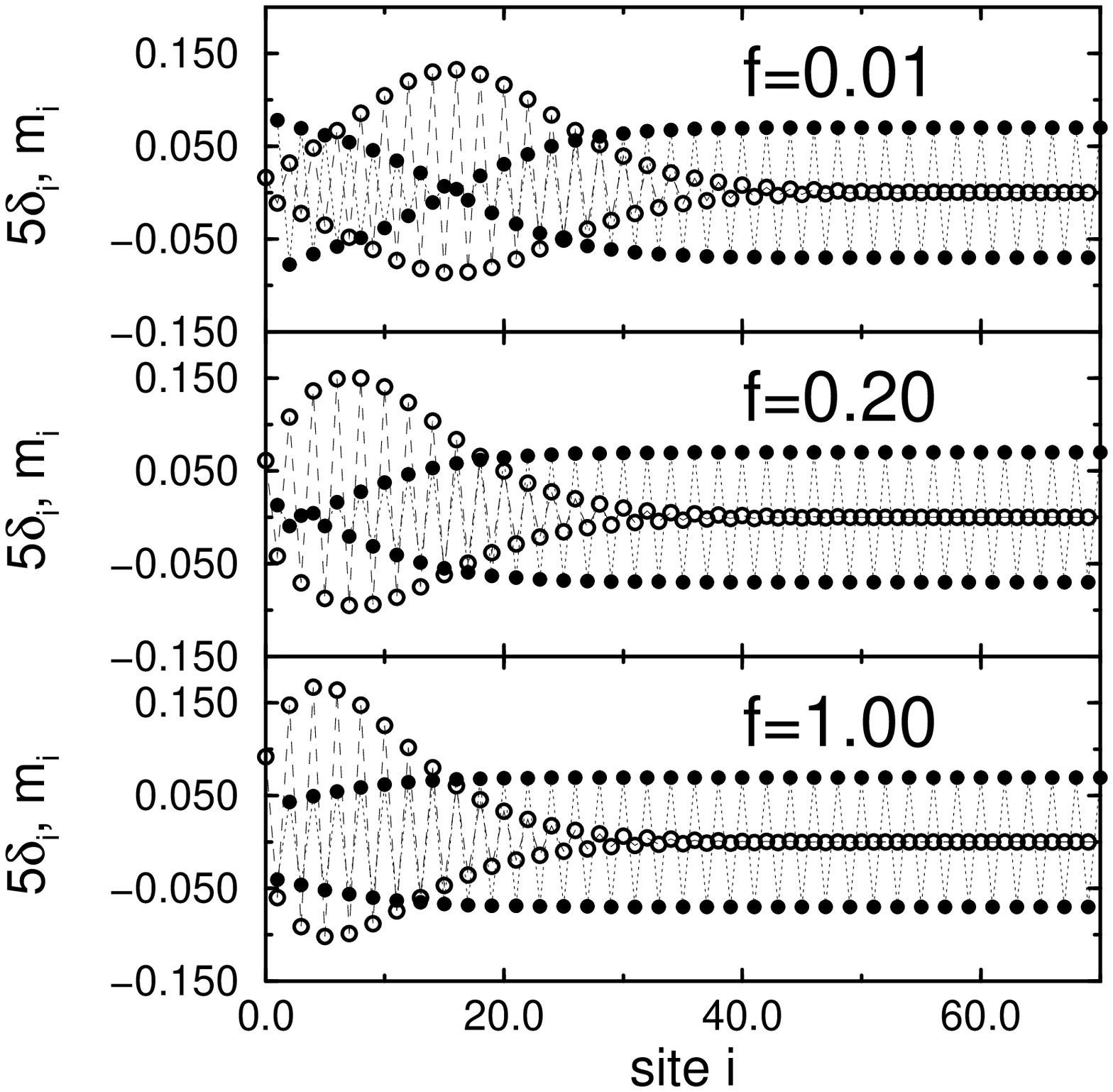,width=6cm}
\caption[fig4]{Local distortions (enhanced) and magnetizations in the
vicinity of an impurity (chain end) at site -1 for $K=18.064$, $\alpha=0.35$
 and $f$ as displayed.}

\label{impur}

\end{figure}

Fig.~\ref{impur} displays the results for various elastic
interchain interactions.
The soliton is {\em not} localized at the impurity but at a certain
distance. For lower $f$ values the soliton resembles very much the ones
in the I phase (cf. Fig.~\ref{inkomm}). On increasing $f$ the soliton 
is squeezed more and more towards the impurity.
Analogous results can be found by QMC \cite{hanse98}, too.

Sure enough, the soliton is bound to the impurity confirming previous
ideas \cite{khoms96,els98}. 
Moreover, the first excitations for the same distortions
and in the same spin sector are found below the singlet-triplet gap at
52\% of $\Delta_{\rm trip}$ 
for $f=0.01$ and at 64\% for $f=1$. This matters for the 
 spectroscopic analysis.

\section{Local Renormalization}
Most remarkable is
the difference between the distortive soliton width $\xi_{\rm d}$ and the
magnetic soliton width $\xi_{\rm m}$ in the numerical 
results (cf. Fig.~\ref{inkomm}). It amounts to 30\% at $\alpha=0.35$
depending mainly on the frustration for low soliton concentrations 
\cite{uhrig99b}.
The challenge is to extend the existing continuum
description to account for this fact. Let us revisit
the semiclassical treatment of the bosonized description of the 
spin-Peierls problem \cite{nakan80}. Minimizing the total
energy the variation of the distortion $\delta(x)$ leads to 
\begin{equation}
\delta(x) \propto e^{-2\sigma}\cos(2\phi_{\rm class})
\end{equation}
where $\sigma:=\langle \hat\phi^2 \rangle$ denotes the 
renormalizing fluctuations
of the local bosonic field $\hat\phi$ about the classical field 
$\phi_{\rm class}$. 
A soliton corresponds to a solution where $\phi_{\rm class}$ increases
by $\pi$ in a kink-like fashion. If $\sigma$ is assumed to take the 
spatially constant value that it has in the ground state \cite{nakan80}
 one obtains
\begin{equation}
\delta(x)/\delta = \cos(2\phi_{\rm class}) = \tanh(x/\xi)\ ,
\end{equation}
wherein $\xi$ is given by the ratio $v_{\rm S}/\Delta_{\rm trip}$ 
of the spin wave velocity and the gap. The alternating component of
the magnetization $a(x)$ is proportional to $\sin(2\phi_{\rm class})$.
Hence, one has $a(x) \propto \sqrt{1-\tanh^2(x/\xi)} = 1/\cosh(x/\xi)$.

\begin{figure}
\vspace*{-5mm}
\hfill\psfig{figure=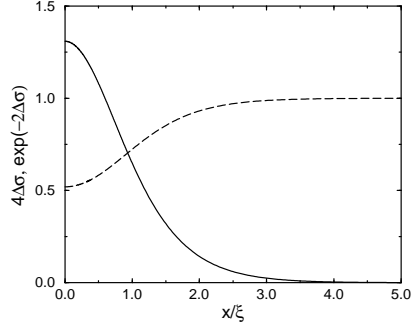,width=6cm}
\vspace*{-5mm}
\caption[fig3]{Local fluctuations as function of $x$. The deviation
$\Delta\sigma$ from the ground state value (solid line) and the 
renormalizing factor (dashed) is shown.}

\label{locren}

\end{figure}

But the presence of the soliton induces a deviation $\Delta\sigma$
from its ground state value. Fig.~\ref{locren} displays a generic
result for this deviation. 
It is calculated on top of the solution
of Nakano and Fukuyama \cite{nakan80}.
The alternating component
\begin{equation}
a(x) \propto \sqrt{1-\exp(4\Delta\sigma)\tanh^2(x/\xi)}
\end{equation}
is thus indeed narrower than before due to the spatial dependence of
the renormalization factor.
The result in Fig.~\ref{locren} is only a first step 
since the influence of the altered magnetic behaviour is not included.
Yet it is clear that the renormalization is a local quantity and that this is
 the origin of the difference between $\xi_{\rm d}$ and $\xi_{\rm m}$.

\section{Phasons}
In Sect.~1 the discrepancy between 
theoretical and experimental amplitudes of the alternating local 
magnetizations was solved by an averaging procedure (\ref{mittel}).
Passing to adaptive modulations does not change the amplitudes much
\cite{uhrig99b}. Hence, one still has to find the microscopic origin for the
averaging.

Non-adiabatic effects are in fact responsible. The soliton lattice
oscillates about its equilibrium positions. These
oscillations are best understood in a continuum description which is
well justified if the typical length $\xi$ is noticeably larger than
the lattice constant. In such a continuum description the modulation
$\delta(x)$ can be shifted along the chains without energy cost. This
continuous translational invariance is spontaneously broken by the 
soliton lattice. Hence, there are massless Goldstone modes, the
so-called phasons. They are  analogous to the phonons of a
crystal lattice except that they do not have three branches
but only with one. While the atoms of a crystal lattice can be shifted
in all three spatial directions the solitons can be shifted only
along the chains.

Albeit the ideal spin-Peierls system is magnetically one-dimensional
the distortions on different chains are elastically coupled. Thus, the 
phasons are governed by a 3D, though anisotropic, dispersion.
The dispersion parameters are determined from the anisotropy of the 
correlation lengths assuming a Ginzburg-Landau description \cite{bhatt98}.
The corresponding $T^3$ term in the specific heat has been measured
\cite{loren96} and theory and experiment agree astonishingly well 
\cite{bhatt98}.

The zero point motion and the excited motion of phasons lead to
the averaging (\ref{mittel}). Let us denote the adiabatic result
for the local magnetizations by 
$m_i = a({r}_i)\cos(\pi {r}_i) + u({r}_i)$ where $r_i$ 
is the component along the chains, $a(r_i)$ the alternating component
and $u(r_i)$ the uniform component of the magnetizations.
A local shift can be implemented by replacing
$\pi {r}_i \to \pi {r}_i+\hat \Theta({\bf r}_i)$ where 
$\hat \Theta({\bf r}_i)$ denotes a  phase shift operator.
 On the long time scales
of a NMR measurement one measures
\begin{equation}
\label{mi-measured}
m_i^{\rm exp} = \langle m_i\rangle
= a({r}_i) \gamma' \cos(\pi {r}_i) + u({r}_i)
\end{equation}
 with
$\gamma' := \exp\left(-\frac{1}{2N}\sum_i
\langle\hat \Theta^2({\bf r}_i)\rangle
\right) < 1$.
So there is an amplitude reduction engendered by the local
fluctuations. The factor $\gamma'$ is comparable to a Debye-Waller factor.

Using the values fixed previously \cite{bhatt98} yields
$\gamma' = 0.16 \exp(- (T/T^*)^2/2)$ with $T^*\approx 16.9$K \cite{uhrig99b}.
Eq. (\ref{mittel}) is retrieved by estimating  $a(r)$ and $u(r)$ from the
discrete values $m_i$ by
$
a({r}_i) = m_i/2-(m_{i-1}+m_{i+1})/4$
and by
$u({r}_i) = m_i/2+(m_{i-1}+m_{i+1})/4$.
Inserting these formulae into Eq. (\ref{mi-measured}) yields 
Eq. (\ref{mittel}) with $\gamma=(1-\gamma')/4$. At $T=0$, $\gamma$
takes the value $0.21$ in accordance with experiment 
\cite{uhrig98a,uhrig99b}.

\section{Conclusions}

In this report the modulated phases of spin-Peierls systems were
discussed. Such modulations are induced either by magnetic field
or by impurities. In both ways the singlet pairing in the 
D phase is broken and spinons are freed. The lattice 
distortion adapts to the spinon by forming a zero to which the spinon
is bound. This new entity constitutes the spin-Peierls soliton.

The order of the transition D  $\to$  I phase on increasing field
depends on model details. Imposed sinusoidal modulation leads
to a pronounced first order transition. Allowing the system
to choose an optimum modulation makes the transition continuous
{\em if} the elastic energy is wave vector independent. If the
elastic energy itself pins the modulation to $\pi$ a weak first
order transition is found.

Doping induced solitons are bound to their generating impurity.
The distortion pattern between impurity and soliton is not coherent
with the bulk pattern. This costs energy which acts as a confining
potential. Binding occurs for which experimental evidence exists 
\cite{hassa98,els98}.

The difference between magnetic and distortive soliton width
could be traced back to the so far neglected spatial dependence
of the renormalizing local fluctuations. A fully self consistent
analysis is in progress.

The reduction of the alternating magnetic amplitude due to phasons
 provides striking evidence for the importance
of the lattice dynamics. The inclusion of non-adiabatic effects
on top of an otherwise adiabatic calculation might still 
be unsatisfactory. So other approaches to non-adiabatic behaviour
should be extended to the I phase \cite{uhrig98b}.

I thank C. Berthier, J.P. Boucher, B. B\"uchner, T. Lorenz, 
M. Horvati\'c,  Th. Nattermann
for fruitful discussions, F. Sch\"onfeld for  reliable 
numerical work, E. M\"uller-Hartmann for generous support and 
 G. G\"untherodt's group for  intense collaboration.
Financial support of the DFG by the SFB 341 is acknowledged. 


\begin{thebibliography}{99}
\bibitem{hase93a}
M. Hase, I. Terasaki, K. Uchinokura, Phys. Rev. Lett. {\bf 70},  3651
  (1993).

\bibitem{bray83}
J.~W. Bray, L.~V. Interrante, I.~C. Jacobs, J.~C. Bonner,  in {\em Extended
  Linear Chain Compounds}, by J.~S. Miller (Plenum Press, New York,
  1983), Vol.~3, p.\ 353;
J.~P. Boucher, L.~P. Regnault, J. Phys. I France {\bf 6},  1939  (1996).

\bibitem{brazo80a}
S.~A. Brazovski\v{\i}, Sov. Phys. JETP {\bf 51},  342  (1980);
S.~A. Brazovski\v{\i}, S.~A. Gordyunin, N.~N. Kirova, JETP Lett. {\bf 31},
  456  (1980).

\bibitem{nakan80}
T. Nakano, H. Fukuyama, J. Phys. Soc. Jpn. {\bf 49},  1679  (1980);
{\it ibid.} {\bf 50},  2489  (1981).

\bibitem{merts81}
J. Mertsching, H.~J. Fischbeck, Phys. Stat. Sol. (b){\bf103},  783
  (1981).

\bibitem{horov81}
B. Horovitz, Phys. Rev. Lett. {\bf 46},  742  (1981);
{\it ibid.} Phys. Rev. B{\bf35},  734  (1987).

\bibitem{buzdi83a}
A.~I. Buzdin, V.~V. Tugushev, Sov. Phys. JETP {\bf 58},  428  (1983);
A.~I. Buzdin, M.~L. Kuli\'c, V.~V. Tugushev, Solid State Commun. {\bf 48},
  483  (1983).

\bibitem{fujit84}
M. Fujita, K. Machida, J. Phys. Soc. Jpn. {\bf 53},  4395  (1984);
{\it ibid.}  J. Phys. C{\bf21},  5813  (1988).

\bibitem{hijma85}
T.~W. Hijmans, H.~B. Brom, L.~J. de~Jongh, Phys. Rev. Lett. {\bf 54},  1714
 (1985).

\bibitem{kiryu95}
V. Kiryukhin, B. Keimer, Phys. Rev. B{\bf52},  704  (1995);
V. Kiryukhin, B. Keimer, J.~P. Hill, A. Vigliante, Phys. Rev. Lett. {\bf
  76},  4608  (1996);
V. Kiryukhin {\it et~al.}, Phys. Rev. B{\bf54},  7269  (1996).

\bibitem{fagot96}
Y. Fagot-Revurat {\it et~al.}, Phys. Rev. Lett. {\bf 77},  1861  (1996);
M. Horvati\'c {\it et~al.}, Physica B{\bf246-247},  22  (1998).

\bibitem{horva99}
M. Horvati\'c {\it et~al.}, cond-mat/9812370.

\bibitem{uhrig99b}
G.~S. Uhrig, F. Sch\"onfeld, J. Boucher, M. Horvati\'c,
  cond-mat/9902272.

\bibitem{uhrig99a}
G.~S. Uhrig, F. Sch\"onfeld, M. Laukamp, E. Dagotto, Eur. Phys. J. B{\bf7},
  67  (1999).

\bibitem{schon98}
F. Sch\"onfeld, G. Bouzerar, G.~S. Uhrig, E. M\"uller-Hartmann, Eur. Phys.
  J. B{\bf5},  521  (1998).

\bibitem{cross79}
M.~C. Cross, D.~S. Fisher, Phys. Rev. B{\bf19},  402  (1979).

\bibitem{mulle81}
G. M\"uller, H. Thomas, H. Beck, J.~C. Bonner, Phys. Rev. B{\bf24},  1429
   (1981).

\bibitem{uhrig98a}
G.~S. Uhrig, F. Sch\"onfeld, J. Boucher, Europhys. Lett. {\bf 41},  431
  (1998).

\bibitem{feigu97}
A. Feiguin, J. Riera, A. Dobry, H. Ceccatto, Phys. Rev. B{\bf56},
  14607  (1997).

\bibitem{forst98}
D. F\"orster, Y. Meurdesoif, B. Malet, cond-mat/9802245.

\bibitem{palme96a}
W. Palme {\it et~al.}, J. Appl. Phys. {\bf 79},  5384  (1996).

\bibitem{ammer97}
U. Ammerahl {\it et~al.}, Z. Phys. B{\bf102},  71  (1997).

\bibitem{loren97a}
T. Lorenz {\it et~al.}, Phys. Rev. B{\bf55},  5914  (1997).

\bibitem{cross79b}
M.~C. Cross, Phys. Rev. B{\bf20},  4606  (1979).

\bibitem{bhatt98}
S.~M. Bhattacharjee, T. Nattermann, C. Ronnewinkel, Phys. Rev. B{\bf58},
  2658  (1998).

\bibitem{loren98}
T. Lorenz {\it et~al.}, Phys. Rev. Lett. {\bf 81},  148  (1998).

\bibitem{zang97}
J. Zang, S. Chakravarty, A.~R. Bishop, Phys. Rev. B{\bf55},  R14705
  (1997).

\bibitem{khoms96}
D. Khomskii, W. Geertsma, M. Mostovoy, Czech. J. Phys. {\bf 46},
  3239  (1996).

\bibitem{marti96b}
G.~B. Martins, E. Dagotto, J.~A. Riera, Phys. Rev. B{\bf54},  16032
  (1996);
G.~B. Martins, M. Laukamp, E. Dagotto, J.~A. Riera, Phys. Rev. Lett. {\bf
  78},  3563  (1997);
M. Laukamp {\it et~al.}, Phys. Rev. B{\bf57},  10755  (1998).

\bibitem{hassa98}
A.~K. Hassan {\it et~al.}, Phys. Rev. Lett. {\bf 80},  1984  (1998).

\bibitem{els98}
G. Els {\it et~al.}, Europhys. Lett. {\bf 43},  463  (1998).

\bibitem{hanse98}
P. Hansen, D. Augier, J. Riera, D. Poilblanc,  cond-mat/9805325;
D. Augier {\it et~al.},  cond-mat/9807265.

\bibitem{loren96}
T. Lorenz, U. Ammerahl, R. Ziemes, B. B\"uchner, Phys. Rev. B{\bf54},
  R15610  (1996).

\bibitem{uhrig98b}
G.~S. Uhrig, Phys. Rev. B{\bf 57},  R14004  (1998);
A. Wei\ss{}e, G. Wellein, H. Fehske,  cond-mat/9901262.
\end{thebibliography}



\end{document}